# Comparing Production Cross Sections for QCD Matter, Higgs Boson, Neutrino with Dark Energy in Accelerating Universe

Tooraj Ghaffary[*]

[*]Department of Science, Shiraz Branch, Islamic Azad University, Shiraz, Iran
E.mail: ghaffary@iaushiraz.ac.ir

**Abstract**
In this research, the production cross sections for QCD matter, neutrino and dark energy due to acceleration of Universe is calculated. To obtain these cross sections, the Universe production cross section is multiplied by the particle or dark energy distribution in accelerating Universe. Also missing cross section for each matter and dark energy due to formation of event horizon, is calculated. It is clear that the cross section of particles produced near event horizon of Universe is much larger for higher acceleration of Universe. This is because as the acceleration of Universe becomes larger, the Unruh temperature becomes larger and the thermal radiations of particles are enhanced. There are different channels for producing Higgs boson in accelerating Universe. Universe maybe decay to quark and gluons, and then these particles interact with each other and Higgs boson is produced. Also, some Higgs boson are emitted directly from event horizon of Universe. Comparing Higgs boson cross sections via different channels, it is observed that at lower acceleration, $a_{Universe} < 2.5\ a_{early\ Universe}$, the Universe will not be able to emit Higgs, but is still able to produce a quark and eventually for $a_{Universe} < 1.5\ a_{early\ Universe}$ the Universe can only emit massless gluons. As the acceleration of Universe at the LHC increases, $a_{Universe} > 2.5\ a_{early\ Universe}$, most of Higgs boson production will be due to Unruh effect near event horizon of Universe. Finally comparing the production cross section for dark energy with particle cross sections, it is found that the cross section for dark energy is dominated by QCD matter, Higgs boson and neutrino. This result is consistent with previous predictions for dark energy cross section.

**Key words**: Accelerating Universe; Higgs boson; QCD matter; Cross Section

## 1. Introduction

Recently some researchers have considered different channels for Higgs production from TeV black holes and by summing over all cross sections for Higgs boson radiation from black holes, the effect of this particle radiation from mini black holes is predicted [1,2]. On the other hand, the most important observational advance in cosmology since the early studies of cosmic expansion in the 1920's, was the dramatic and unexpected discovery in the waning years of the twentieth century that the Universe is accelerating. This was first announced in February 1998, based on the concordance of two groups' data on Supernovae Type IA [3].

According the Unruh idea, there is a similar effect in flat space for a detector undergoing the acceleration of Universe to a detector near event horizon of black hole [4]. Following Hawking's discovery that asymptotically-flat stationary black holes emit thermal radiation, the detector in accelerating Universe appears to be in a heat bath at the Unruh temperature, ($T_U = \frac{\hbar a_{Universe}}{2\pi k_B c}$ where $a_{Universe}$ is the acceleration of Universe) [5,6].

The acceleration of Universe may change the way we search for new particles. Decays of Universe, tagged with prompt leptons or photons, offer low-background environment for searches of new particles. For example, a signature of 125 GeV SM-like Higgs boson can be observed due to acceleration of Universe. As the Unruh temperatures of the accelerating Universe is very high, the Universe can be a source for particle production via Unruh radiation. In fact there can be an enormous amount of particle production due to acceleration of Universe, much more than expected until now.

Recent astrophysical data from distant IA supernovae, Large Scale Structure and Cosmic Microwave Background observations, [3,7,8], show that Universe is accelerating due to some kind of negative-pressure form of matter known as dark energy [9,10]. The simplest candidate for dark energy is the cosmological constant [10], conventionally associated with the energy of the vacuum with constant energy density and pressure. The present observational data favor an equation of state for the dark energy with parameter very close to that of the cosmological constant. The next simple model proposed for dark energy is the quintessence a dynamical scalar field which slowly rolls down in a potential. Many theoretical attempts toward reconstructing the potential and dynamics of the scalar fields have been done in the literature by establishing a connection between holographic energy density and a scalar-field model of dark energy [11]. However, in the present research, to obtain dark energy cross section, the Universe production cross section is multiplied by thermal distribution of dark energy. The dark energy states and missing dark energy cross section are another important subject that investigated in present research.

In this paper the effect of Higgs boson radiation due to acceleration of Universe on hadronic cross section at LHC will be investigated. So in section (2) the production cross section for QCD matter due to acceleration of Universe is obtained. In section (3), different channels for producing Higgs boson in accelerating Universe is considered. In section (4), the radiation of neutrino from event horizon of Universe is studied. Finally, in section (5), the production cross section for dark energy is calculated

## 2. The Production Cross Section for QCD Matter Due to Acceleration of Universe

In this part the results of the derivation of Hawking radiation for Scalar, [11], and Dirac fields, [13], in black holes into the quarks and gluons in accelerating Universe are extended and the QCD matter states in accelerating Universe are found. Then the thermal distributions of QCD matter states inside and outside the event horizon of Universe are calculated.

Using the transverse free field operator, the gluon quantization is, [14]:

$$A_\mu^a = \int \frac{d^3k}{(2\pi)^3\sqrt{2\omega(k)}} \sum_{\lambda=1}^{3} \varepsilon_\mu^\lambda \left[ a_\lambda^a(k)e^{-ikx} + a_\lambda^{a\dagger}(k)e^{ikx} \right] \quad (1)$$

where $\varepsilon_\mu^\lambda$ denoted as polarization vectors that satisfies the transversally condition:

$$k^\mu \varepsilon_\mu^\lambda = 0 \quad (2)$$

It can be considered that the gluon field satisfies the wave equation [15]:

$$(-g)^{\frac{1}{2}} \frac{\partial}{\partial x^\mu} \left[ g^{\mu\nu}(-g)^{\frac{1}{2}} \frac{\partial}{\partial x^\nu} \right] A_{\rho,s}^a = 0 \quad (3)$$

where the upper index, ($a = 1,...8$), is due to the eight color of gluons, the lower index, "s", (s=1,s$_z$=+1,-1), points to the spin of gluon, ρ, (ρ=1,…d), denotes the vector index, "d" and $g^{\mu\nu}$ represented as the number of dimensions the metric tensor respectively. In order to get the Unruh state for gluons, the Eq. (3) is solved in Kruskal space-time. In Kruskal coordinates the metric of the accelerating Universe becomes [16]:

$$ds^2 = -2\frac{e^{-\frac{a_{Universe}\, r}{2}}}{a_{Universe}\, r} d\bar{u}d\bar{v} + r^2 d\Omega^2 + g_{ij}dx^i dx^j \tag{4}$$

$$\bar{u} = -\frac{4}{a_{Universe}} e^{-\frac{a_{Universe}\, u}{4}}, \quad \bar{v} = -\frac{4}{a_{Universe}} e^{-\frac{a_{Universe}\, v}{4}}$$

$$u = t - r^*, \; v = t + r^*, \; r^* = -r - \frac{2}{a_{Universe}} \ln\left| r - \frac{2}{a_{Universe}} \right| \tag{5}$$

where $a_{Universe}$ is the acceleration of Universe. The Kruskal extension of the Universe space-time is shown in Fig. 1 [16,17].

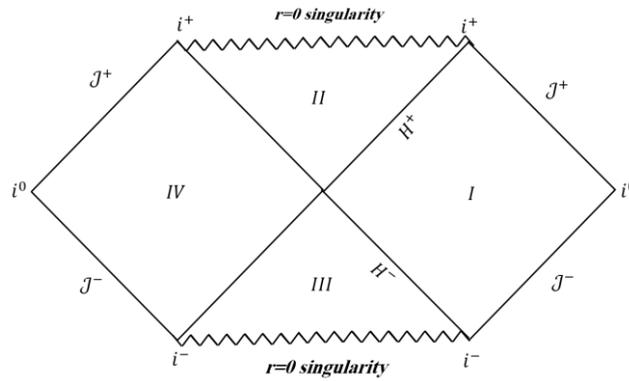

**Figure 1.** The Kruskal extension of the Universe space-time similar to black hole space-time [16,17]. In region *I*, null asymptotes $H^+$ and $H^-$ act as future and past event horizons, respectively. The boundary lines labelled $J^+$ and $J^-$ are future and past null infinities, respectively, and $i^0$ is the space-like infinity [12].

Considering the fact that $\bar{v} = 0$ on $H^-$, [12], the original positive frequency normal mode on past horizon can be estimated as:

$$A^a_{\mu,s} \propto \varepsilon^a_{\mu,s}(e^{-i\omega u}) = \varepsilon^a_{\mu,s} \left( \frac{a_{Universe}\, |\bar{u}|}{4} \right)^{-\frac{i4\omega}{a_{Universe}}}$$

$$= \begin{cases} \varepsilon^a_{\mu,s} \left( \dfrac{-a_{Universe}\, \bar{u}}{4} \right)^{-\frac{i4\omega}{a_{Universe}}} & (region\; I) \\[1em] \varepsilon^a_{\mu,s} \left( \dfrac{a_{Universe}\, \bar{u}}{4} \right)^{-\frac{i4\omega}{a_{Universe}}} & (region\; II) \end{cases} \tag{6}$$

where $\omega$ supposes as the gluon energy. In Eq. (6), it can be used the fact that $(-1)^{-\frac{i4\omega}{a_{Universe}}} = e^{\frac{4\pi\omega}{a_{Universe}}}$. From Eq. (6), it is observed that the states in the horizon satisfy the following condition [1]:

$$\left(A^a_{\mu,s_{in}} - e^{-\frac{4\pi\omega}{a_{Universe}}} A^a_{\mu,s_{out}}\right)|Universe, \mu, a, s\rangle_{in\otimes out} = 0 \tag{7}$$

or we have:

$$\left(A^a_{\mu,s_{out}} - \tanh r_\omega A^a_{\mu,s_{in}}\right)|Universe, \mu, a, s\rangle_{in\otimes out} = 0$$
$$\text{with } \tanh r_\omega = e^{\frac{-4\pi\omega}{a_{Universe}}} \tag{8}$$

Using the expansion in modes for gluons (Eq. (1)) we will have:

$$\left(\alpha^a_{\lambda,s_{out}} - \tanh r_\omega \alpha^{a\dagger}_{\lambda,s_{in}}\right)|Universe, \lambda, a, s\rangle_{in\otimes out} = 0 \quad for\ n \neq 0$$
$$\tanh r_\omega = e^{\frac{-4\pi\omega}{a_{Universe}}} \tag{9}$$

Now, we suppose that the Kruskal vacuum, $|Universe, \lambda, a, s\rangle_{in\otimes out}$, is linked to the Schwarzschild vacuum $|0\rangle_S$ by

$$|Universe, \lambda, a, s\rangle_{in\otimes out} = F\left(\alpha^a_{\lambda,s_{out}}, \alpha^a_{\lambda,s_{in}}\right)|0\rangle_S \tag{10}$$

where $F$ is some function to be determined later.

From $\left[\alpha^a_{\lambda,s_{out}}, \alpha^a_{\lambda,s_{in}}\right] = 1$, we obtain $\left[\alpha^a_{\lambda,s_{out}}, \left(\alpha^{a\dagger}_{\lambda,s_{in}}\right)^m\right] = \frac{\partial}{\partial \alpha^{a\dagger}_{\lambda,s_{out}}}\left(\alpha^{a\dagger}_{\lambda,s_{in}}\right)^m$ and $\left[\alpha^a_{\lambda,s_{out}}, F\right] = \frac{\partial F}{\partial \alpha^{a\dagger}_{\lambda,s_{out}}}$. Using Eqs. (9) and (10), we will arrive to the following differential equation for $F$.

$$\left(\frac{\partial F_\rho}{\partial \alpha^{a\dagger}_{\lambda,s_{out}}} - \tanh r_\omega \alpha^{a\dagger}_{\lambda,s_{in}} F\right) = 0 \tag{11}$$

and the solution will be:

$$F = e^{\tanh r_\omega \alpha^{a\dagger}_{\lambda,s_{out}} \alpha^{a\dagger}_{\lambda,s_{in}}} \tag{12}$$

Using Eq. (12) and (10) and by properly normalizing the state vector, we have:

$$|Universe, \lambda, a, s\rangle_{in \otimes out} = e^{\tanh r_\omega \alpha^{a^\dagger}_{\lambda,s_{out}} \alpha^{a^\dagger}_{\lambda,s_{in}}} |0\rangle_s$$

$$= \frac{1}{\cosh r_\omega} \sum_n (\tanh r_\omega)^n |n, \lambda, a, s\rangle_{in} \otimes |n, \lambda, a, s\rangle_{out} \quad (13)$$

In order to obtain the state vector, we sum over transversal (physical) degrees of freedom:

$$|Universe, \mu, a, s\rangle_{in \otimes out} = \sum_{\lambda=1}^{3} \varepsilon^\lambda_\mu |Universe, \lambda, a, s\rangle_{in \otimes out}$$

$$= \frac{1}{\cosh r_\omega} \sum_n (\tanh r_\omega)^n |n, \mu, a, s\rangle_{in} \otimes |n, \mu, a, s\rangle_{out} \quad (14)$$

where $|n, \mu, a, s\rangle_{in}$ and $|n, \mu, a, s\rangle_{out}$ denote as orthonormal bases for $H_{in}$ and $H_{out}$ respectively. It is observed that the ground state for gluons near Universe horizon is a maximally entangled with two-mode squeezed states on outside and inside Hilbert spaces of Universe. As one can see from Eq. (14), the production of different number of gluons is happened with different probabilities inside and outside of Universe. These probabilities are related to acceleration of Universe and the gluon's energy. Thermal distribution for these groups of gluons inside the Universe is derived as follows [1]:

$$N^{gluon}_{\omega,color,spin,in} = {}_{in \otimes out}\langle Universe, \mu, a, s| \alpha^{a^\dagger}_{\mu,s_{in}} \alpha^{a}_{\mu,s_{in}} |Universe, \mu, a, s\rangle_{out \otimes in}$$

$$= {}_{out}\langle n, \mu, a, s| {}_{in}\langle n, \mu, a, s| \frac{1}{\cosh^2(r_\omega)} \alpha^{a^\dagger}_{\mu,s_{in}} \alpha^{a}_{\mu,s_{in}} \sum_{n=0}^{\infty} \tanh^{2n}(r_\omega) |n, \mu, a, s\rangle_{in} |n, \mu, a, s\rangle_{out}$$

$$= {}_{out}\langle n, \mu, a, s| {}_{in}\langle n-1, \mu, a, s| \frac{1}{\cosh^2(r_\omega)} \sum_{n=0}^{\infty} \tanh^{2n}(r_\omega) (n)|n-1, \mu, a, s\rangle_{in} |n, \mu, a, s\rangle_{out} \quad (15)$$

$$= \frac{1}{\cosh^2(r_\omega)} \sum_{n=0}^{\infty} e^{\frac{-8\pi\omega}{a_{Universe}}} (n) = \frac{1}{\cosh^2(r_\omega)} \frac{e^{\frac{-8\pi\omega}{a_{Universe}}}}{\left(1 - e^{\frac{-8\pi\omega}{a_{Universe}}}\right)^2} = \frac{e^{\frac{-8\pi\omega}{a_{Universe}}}}{1 - e^{\frac{-8\pi\omega}{a_{Universe}}}}$$

where $\alpha^{a^\dagger}_{\mu,s_{in}}$ and $\alpha^{a}_{\mu,s_{in}}$ are creation and annihilation operators that operate on Universe inside states of gluons. $N^{gluon}_{\omega,color,spin,in}$ introduces as the thermal distribution for gluons with energy $\omega$ and with a one special color and spin inside the accelerating Universe. We usually detect this distribution and miss the thermal distribution outside the event horizon of Universe. That is because we are located inside the Universe. This is in contradiction with black holes problem. In black holes, we are outer observer and detect the thermal distribution outside the event horizon. We can easily calculate the outer distribution as following [1]:

$$N_{\omega,color,spin,out}^{gluon} = {}_{in\otimes out}\langle Universe, \mu, a, s | \alpha_{\mu,s_{out}}^{a\dagger} \alpha_{\mu,s_{out}}^{a} | Universe, \mu, a, s\rangle_{out\otimes in}$$

$$= {}_{out}\langle n, \mu, a, s | {}_{in}\langle n, \mu, a, s | \frac{1}{\cosh^2(r_\omega)} \alpha_{\mu,s_{out}}^{a\dagger} \alpha_{\mu,s_{out}}^{a} \sum_{n=0}^{\infty} \tanh^{2n}(r_\omega) |n, \mu, a, s\rangle_{in} |n, \mu, a, s\rangle_{out}$$

$$= {}_{out}\langle n, \mu, a, s | {}_{in}\langle n-1, \mu, a, s | \frac{1}{\cosh^2(r_\omega)} \sum_{n=0}^{\infty} \tanh^{2n}(r_\omega) (n) |n-1, \mu, a, s\rangle_{in} |n, \mu, a, s\rangle_{out} \quad (16)$$

$$= \frac{1}{\cosh^2(r_\omega)} \sum_{n=0}^{\infty} e^{\frac{-8\pi\omega}{a_{Universe}}} (n) = \frac{1}{\cosh^2(r_\omega)} \frac{e^{\frac{-8\pi\omega}{a_{Universe}}}}{\left(1 - e^{\frac{-8\pi\omega}{a_{Universe}}}\right)^2} = \frac{e^{\frac{-8\pi\omega}{a_{Universe}}}}{1 - e^{\frac{-8\pi\omega}{a_{Universe}}}}$$

where $\alpha_{\mu,s_{out}}^{a\dagger}$ and $\alpha_{\mu,s_{out}}^{a}$ are creation and annihilation operators that operate on Universe outside states of gluons. $N_{\omega,color,spin,out}^{gluon}$ is introduced as the thermal distribution for gluons with energy $\omega$ and with a one special color and spin outside the accelerating Universe. Near event horizon of Universe, in order to obtain the total gluon cross section produced due to acceleration of Universe, it is needed to multiply the Universe production cross section by the number of gluons produced near an event horizon of Universe [1].

$$\sigma_{Universe \to g} = \int d\omega\, N_{\omega,color,spin,in}^{gluon} \sigma_{Universe}$$

$$= \frac{1}{\sigma_{early\,Universe}^2} \left[\frac{a_{Universe}}{a_{early\,Universe}} \left(\frac{4\Gamma(\frac{7}{2})}{3}\right)\right]^2 \quad (17)$$

$$\times \frac{a_{Universe}}{8\pi} \left[\ln(a_{Universe} M_g) + \ln(a_{Universe}^2 M_g^2 - 1) + 2\ln(2)\right]$$

In which we can calculate the production cross section for Universe in comparing with black hole production at LHC, [18,19]:

$$\sigma_{Universe} = \frac{1}{a_{early\,Universe}^2} \left[\frac{a_{Universe}}{a_{early\,Universe}} \left(\frac{8\Gamma(\frac{d+3}{2})}{d+2}\right)\right]^{\frac{2}{d+1}} \quad (18)$$

where $a_{Universe}$ and $a_{early\,Universe}$ are the acceleration of Universe and acceleration of early Universe respectively. "d" is the number of extra dimensions. The missing cross section for gluon can be obtained as follow:

$$\sigma_{missing, Universe \to g} = \int d\omega\, N_{\omega,color,spin,out}^{gluon} \sigma_{Universe} = \sigma_{Universe \to g} \quad (19)$$

We can't access to these particles due to formation of event horizon in accelerating Universe.

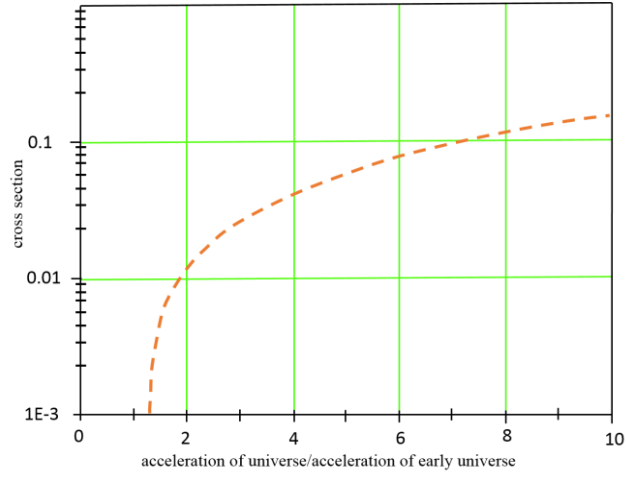

**Figure 2.** The gluon production cross section as a function of acceleration of Universe.

In Fig.2 we present the gluon production cross section due to acceleration of Universe as a function of acceleration of Universe. In this plot early acceleration of Universe is normalized to one and all accelerations are compared with it. The acceleration of Universe is given by [11]:

$$a_{Universe} = t^{\frac{2}{3k}} \quad , \quad a_{early\ Universe} = t_0^{\frac{2}{3k}} \tag{20}$$

where "k" is the constant parameter and t and $t_0$ are present time and early time respectively. We can write:

$$\frac{a_{Universe}}{a_{early\ Universe}} = \left(\frac{t}{t_0}\right)^{\frac{2}{3k}} = N \rightarrow a_{Universe} = N a_{early\ Universe}, \quad \text{(N isn't integer)} \tag{21}$$

It is clear that the cross section of gluon produced near an event horizon of Universe is much larger for higher acceleration of Universe. This is because as the acceleration of Universe becomes larger, the Unruh temperature becomes larger and the thermal radiation of gluon is enhanced.

Now quark production via Unruh effect in accelerating Universe is investigated. In accelerating Universe, the quark equation can be derived as [13]:

$$[i\gamma^\mu(\partial_\mu + \Gamma_\mu) - m_q]\psi_{q,s}^a = 0 \tag{22}$$

where $a = 1, 2, 3$ is concerned with the color of quark, "s" points to the quark spin, "$m_q$" is the quark mass and q, (u,d,s,c,t,b), are represented as the quark flavors. The affine connection is defined by:

$$\Gamma_\mu = -\frac{1}{4} \gamma^\nu (\partial_\mu \gamma^\nu + \Gamma^\nu_{\mu\lambda} \gamma^\nu) \tag{23}$$

The gamma matrices are given by $\gamma_\mu = e_\mu^a \bar{\gamma}_a$ where $e_\mu^a$ are tetrads, $\bar{\gamma}_a$ are the gamma matrices for the inertial frame, [20], and the Levi-Civita connection coefficients $\Gamma^\nu_{\mu\lambda}$ can be obtained by the Lagrange method [13]. Solution for Eq. (22), Using the Kruskal coordinate, is:

$$\psi^a_{q,s_z} \propto \begin{cases} u^{+a}_{r,q} \left( \dfrac{-a_{Universe}\, \bar{u}}{4} \right)^{-\frac{i4\omega}{a_{Universe}}} & (region\ I) \\ \\ u^{-\bar{a}}_{r,\bar{q}} \left( \dfrac{a_{Universe}\, \bar{u}}{4} \right)^{-\frac{i4\omega}{a_{Universe}}} & (region\ II) \end{cases} \tag{24}$$

where $u^{+a}_{r,q}$ and $u^{-\bar{a}}_{r,\bar{q}}$ refer to the quark and antiquark spin wave functions respectively. Considering calculations in ref. [13], we have:

$$|Universe, q, a, s\rangle_{in \otimes out} = \cos(r_\omega) \sum_{n=0,1} (\tanh r_\omega)^n\, |n, q, a, s\rangle_{in} \otimes |n, \bar{q}, \bar{a}, s\rangle_{out}$$

$$\tanh r_\omega = e^{\frac{-4\pi\omega}{a_{Universe}}}, \qquad \cos(r_\omega) = \left( 1 + e^{\frac{-4\pi\omega}{a_{Universe}}} \right)^{-\frac{1}{2}} \tag{25}$$

where $|n, q, a, s\rangle_{in}$ and $|n, \bar{q}, \bar{a}, s\rangle_{out}$ denote quark and antiquark states inside and outside of event horizon of accelerating Universe. The thermal distribution for this quark production inside the accelerating Universe is given by:

$$\begin{aligned} N^{quark}_{\omega,color,s,in} &= {}_{in \otimes out}\langle Universe, q, a, s| c^{a\dagger}_{q,s,in} c^{a}_{q,s,in} |Universe, q, a, s\rangle_{out \otimes in} \\ &= {}_{out}\langle 1, \bar{q}, \bar{a}, s|\, {}_{in}\langle 0, q, a, s| \sin^2(r_\omega) |0, q, a, s\rangle_{in} |1, \bar{q}, \bar{a}, s\rangle_{out} \\ &= \sin^2(r_\omega) = \dfrac{e^{\frac{-8\pi\omega}{a_{Universe}}}}{1 + e^{\frac{-8\pi\omega}{a_{Universe}}}} \end{aligned} \tag{26}$$

where $c^{a\dagger}_{q,s,in}$ and $c^{a}_{q,s,in}$ are creation and annihilation operators that operate on Universe inside states of quarks. $N^{quark}_{\omega,color,s,in}$ is introduced as the thermal distribution for quarks with energy $\omega$ and a special color and spin inside the

Universe. Missing thermal distribution outside the event horizon of Universe can be calculated as following:

$$\begin{aligned}
N^{quark}_{\omega,color,s,out} &= {}_{in\otimes out}\langle Universe, q, a, s | c^{a\dagger}_{q,s,out} c^{a}_{q,s,out} | Universe, q, a, s\rangle_{out\otimes in} \\
&= {}_{out}\langle 0, \bar{q}, \bar{a}, s| \; {}_{in}\langle 1, q, a, s| \sin^2(r_\omega) |1, q, a, s\rangle_{in} |0, \bar{q}, \bar{a}, s\rangle_{out} \\
&= \sin^2(r_\omega) = \frac{e^{\frac{-8\pi\omega}{a_{Universe}}}}{1 + e^{\frac{-8\pi\omega}{a_{Universe}}}}
\end{aligned} \quad (27)$$

where $c^{a\dagger}_{q,s,out}$ and $c^{a}_{q,s,out}$ are creation and annihilation operators that act on Universe outside states of quarks. $N^{quark}_{\omega,color,s,out}$ is introduced as the thermal distribution for quarks with energy $\omega$ and a special color and spin outside the Universe. As mention before, in black hole problem we detect this distribution as an observer outside the event horizon of black hole. However in accelerating Universe, we miss this distribution. The total quark cross section near this event horizon is:

$$\begin{aligned}
\sigma_{Universe \to q} &= \int d\omega \, N^{quark}_{\omega,color,spin,in} \sigma_{Universe} \\
&= \frac{1}{\sigma^2_{early\,Universe}} \left[ \frac{a_{Universe}}{a_{early\,Universe}} \left( \frac{4\Gamma\left(\frac{7}{2}\right)}{3} \right) \right]^2 \\
&\quad \times \frac{a_{Universe}}{8\pi} \left[ \ln(a_{Universe} M_q) + \ln(a_{Universe}^2 M_q^2 - 1) + 2\ln(2) \right]
\end{aligned} \quad (28)$$

and the missing cross section can be calculated as following:

$$\sigma_{missing,Universe \to q} = \int d\omega \, N^{quark}_{\omega,color,spin,out} \sigma_{Universe} = \sigma_{Universe \to q} \quad (29)$$

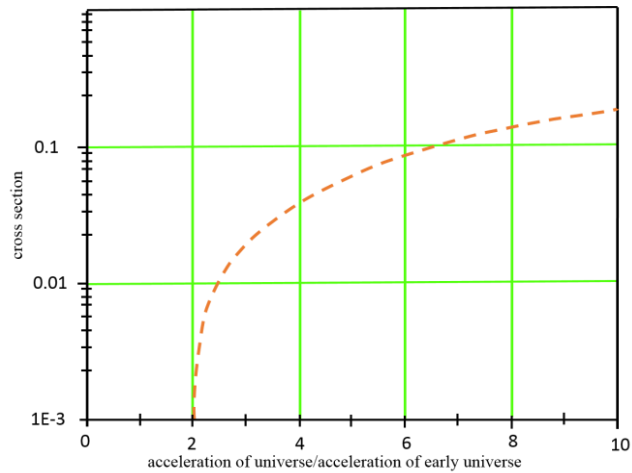

**Figure 3.** The quark production cross section as a function of acceleration of Universe.

In Fig. 3 we present the results for the cross section of quark produced via near event horizon of Universe as a function of acceleration of Universe. In this plot we choose $M_b = 5\ GeV$, $M_H = 125\ GeV$ and $a_{Universe}$ is normalized to one and all accelerations are compared with this acceleration. As can be seen from the Fig. 3, the cross sections increases rapidly when the acceleration increases.

**3. Higgs Production Near Event Horizon of Accelerating Universe**

Colorful black holes at LHC radiate a large amount of QCD matter and less amount of other fermions and bosons. This leads to the production of Higgs bosons via quark and gluon interactions. For this reason we discuss Higgs boson production via QCD matter interactions and give out other channels for its production.

In this section it is shown that Universe decay to a large amount of particles like Higgs boson, quarks and gluons. In fact the Universe is expected to be a particle factory. Because quarks and gluons' energies are very high, these particles are considered as free particles inside the Universe and also, as it is said before, quark and antiquark that are produced inside and outside of event horizon have no interaction with each other. Thus they aren't expected to hadronize unless forming Higgs boson. Also the calculations are shown that some Higgs bosons will be radiated directly from event horizon of Universe. In this research, the different channels for Higgs boson production are studied. Thus the process that Higgs boson are produced via heavy quark and high energy gluon interactions are considered.

First, Universe decays to gluons. Next, gluons interact with each other and Higgs boson is produced. The rate for Higgs production due to interaction between gluons is given by [22]:

$$\Gamma(gg \to H) = \left(\frac{\alpha_s}{\pi}\right)^2 \frac{\pi}{288\sqrt{2}} \left[\frac{3}{2\Lambda}\left(1 + \left(1 - \frac{1}{\Lambda}\right) arcsin^2(\sqrt{\Lambda})\right)\right]^2 \tag{30}$$

where $\alpha_s$ denotes as the strong coupling constant, $\Lambda = \frac{M_H^2}{4M_t^2}$ is the renormalization scale, $M_H$ and $M_t$ are represented as the Higgs boson and the top quark masses respectively.

It is assumed gluons are annihilated and heavy quark - antiquarks with a mass $M_q$ are produced and a subsequent decay to Higgs boson is happened. Here we assume that these heavy quarks are top quarks. Since the top quark mass, ($M_t = 175\ GeV$), is close to Higgs boson mass, these quarks could easily interact with each other and produce Higgs bosons. This Higgs boson production cross section is denoted by $\sigma_H^1$ and calculated as follow [1]:

$$\begin{aligned}
&\sigma_H^1(Universe \to H)\\
&= \int dM_{Universe} \int \frac{d\omega_1}{2\pi} \int \frac{d\omega_2}{2\pi} \int dz\ 2\pi\delta(zM_{Universe} - \omega_1 - \omega_2)\delta\left(M_{Universe} - \frac{1}{a_{Universe}}\right)\\
&\quad \times 2\pi\delta(M_{Universe} - \omega_1 - \omega_2) L^{Universe \to gg}(\omega_1, \omega_2) p_g(\omega_1) p_g(\omega_2) L^{gg \to H}\\
&= \int \frac{d\omega_1}{2\pi} \int \frac{d\omega_2}{2\pi} \int dz\ 2\pi\delta\left(z\frac{1}{a_{Universe}} - \omega_1 - \omega_2\right) 2\pi\delta(M_H - \omega_1 - \omega_2)\\
&\quad \times \frac{e^{\frac{-8\pi\omega_1}{a_{Universe}}}}{1 - e^{\frac{-8\pi\omega_1}{a_{Universe}}}} \frac{e^{\frac{-8\pi\omega_2}{a_{Universe}}}}{1 - e^{\frac{-8\pi\omega_2}{a_{Universe}}}} \frac{1}{a_{early\ Universe}^2} \left[\frac{z}{a_{early\ Universe}(\omega_1 + \omega_2)}\left(\frac{4\Gamma\left(\frac{7}{2}\right)}{3}\right)\right]^2 \frac{1}{\omega_1^2}\frac{1}{\omega_2^2}\\
&\quad \times \sum_{c=1}^{8}(T^cT^c)_{mn}\left(\frac{\alpha_s}{\pi}\right)^2 \frac{\pi}{288\sqrt{2}}\left[\frac{6M_t^2}{(\omega_1+\omega_2)^2}\left(1 + \left(1 - \frac{4M_q^2}{(\omega_1+\omega_2)^2}\right)arcsin^2\left(\sqrt{\frac{(\omega_1+\omega_2)^2}{4M_t^2}}\right)\right)\right]^2 \quad (31)\\
&\approx C_F \delta_{mn}\left(\frac{\alpha_s}{\pi}\right)^2 \frac{\pi}{288\sqrt{2}}\left[\frac{6M_t^2}{M_H^2}\left(1 + \left(1 - \frac{4M_t^2}{M_H^2}\right)arcsin^2\left(\sqrt{\frac{M_H^2}{4M_t^2}}\right)\right)\right]^2\\
&\quad \times \frac{1}{a_{early\ Universe}^2}\left[\frac{a_{Universe}}{a_{early\ Universe}}\left(\frac{4\Gamma\left(\frac{7}{2}\right)}{3}\right)\right]^2 \frac{e^{\frac{-8\pi}{a_{Universe}^2}}}{1 - e^{\frac{-8\pi}{a_{Universe}^2}}}\left[\frac{1}{M_t^2} - \frac{1}{M_H^2} + \ln\left(\frac{M_t}{M_H}\right)\right]
\end{aligned}$$

where $M_{Universe}$ is the Universe mass and "z" is the fraction of Universe energy that transforms to gluon energy. If we assume that Universe is a black hole that we are located inside it and compare the Hawking temperature and Unruh temperature, we conclude that Universe mass is proportional to the inverse of acceleration. When Universe decays to gluons, each of these particles carry a fraction of Universe energy. This means that Universe' energy decreases with each gluon emitted and we should integrate over z. We also define the gluon part, Higgs part and gluon propagator for this integral as:

$$L^{Universe \to gg} = \frac{e^{\frac{-8\pi\omega_1}{a_{Universe}}}}{1-e^{\frac{-8\pi\omega_1}{a_{Universe}}}} \frac{e^{\frac{-8\pi\omega_2}{a_{Universe}}}}{1-e^{\frac{-8\pi\omega_2}{a_{Universe}}}} \frac{1}{a_{early\,Universe}^2} \left[\frac{z}{a_{early\,Universe}(\omega_1+\omega_2)}\left(\frac{4\Gamma(\frac{7}{2})}{3}\right)\right]^2 \quad (32)$$

$$L^{gg \to H} = \sum_{c=1}^{8}(T^c T^c)_{mn}\left(\frac{\alpha_s}{\pi}\right)^2 \frac{\pi}{288\sqrt{2}}\left[\frac{6M_t^2}{(\omega_1+\omega_2)^2}\left(1+\left(1-\frac{4M_q^2}{(\omega_1+\omega_2)^2}\right)arcsin^2\left(\sqrt{\frac{(\omega_1+\omega_2)^2}{4M_t^2}}\right)\right)\right]^2 \quad (33)$$

$$= C_F \delta_{mn}\left(\frac{\alpha_s}{\pi}\right)^2 \frac{\pi}{288\sqrt{2}}\left[\frac{6M_t^2}{(\omega_1+\omega_2)^2}\left(1+\left(1-\frac{4M_t^2}{(\omega_1+\omega_2)^2}\right)arcsin^2\left(\sqrt{\frac{(\omega_1+\omega_2)^2}{4M_t^2}}\right)\right)\right]^2$$

$$C_F = \sum_{c=1}^{8}(T^c T^c)_{mn} = \frac{4}{3}\delta_{mn} \quad for\ SU(3)$$
$$p_g(\omega_1) = \frac{1}{\omega_1^2} \quad (34)$$

where $C_F$ is the color factor, $T^c$'s are color group generators, "*m*" and "*n*" are the color indices for gluons. In order to form a color singlet and hence Higgs boson, two out coming gluons of Universe must have the correct color. For this, one suppression color factor, ($C_F$), should be added to Eq. (33).

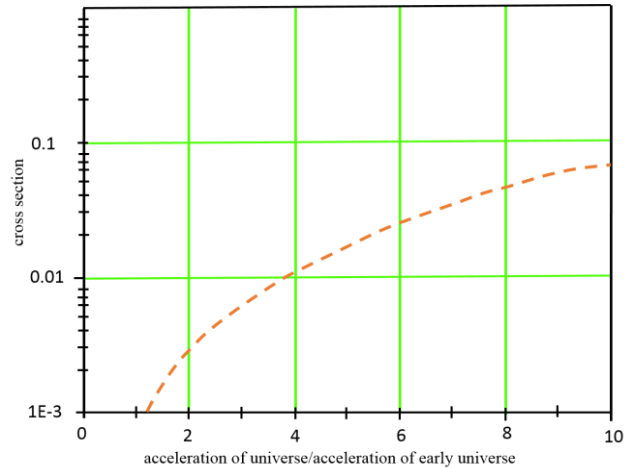

**Figure 4.** The Higgs boson production cross section, ($\sigma(Universe \to gg \to Higgs)$), as a function of acceleration of Universe.

In Fig.4 we present the Higgs boson production cross section via gluon interaction as a function of acceleration of Universe. In this plot we choose $M_t = 175\ GeV$, $M_H = 125\ GeV$, $a_{early\ Universe}$ is normalized to one and all accelerations are compared with this acceleration. Evidently the cross section of Higgs boson produced via gluon fusion near an event horizon of Universe is much larger for higher acceleration of Universe. Since as the acceleration of Universe becomes larger, the thermal radiation of gluon and consequently the Higgs boson production due to gluon interaction is enhanced.

Now the Higgs boson cross section via quark fusion near event horizon of Universe is discussed. The bottom quarks produced near event horizon of Universe can interact with each other and produce Higgs boson. The rate for this interaction is [23]:

$$\Gamma^{b\bar{b}\to H} = \frac{M_b{}^2 \alpha_s}{6v^2 M_H{}^2}\left[-4C_F \ln\left(1-\frac{M_b{}^2}{M_H{}^2}\right) + 2C_F \ln\left(\frac{M_b{}^2}{M_H{}^2}\right) + 2C_F\right] \quad (35)$$

where $v = 246\ GeV$ and $C_F = \frac{4}{3}$

This Higgs boson production cross section via process $Universe \to b\bar{b} \to H$ is denoted by $\sigma_H^2$ [1]:

$$\begin{aligned}
\sigma_H^2(pp \to H) &= \int dM_{Universe} \int \frac{d\omega_1}{2\pi} \int \frac{d\omega_2}{2\pi} \int dz\, 2\pi\delta(zM_{Universe} - \omega_1 - \omega_2)\delta\left(M_{Universe} - \frac{1}{a_{Universe}}\right) \times 2\pi\delta(M_H - \omega_1 - \omega_2) L^{Universe\to q\bar{q}}(\omega_1, \omega_2) p_q(\omega_1) p_{\bar{q}}(\omega_2) L^{q\bar{q}\to H} \\
&= \int \frac{d\omega_1}{2\pi} \int \frac{d\omega_2}{2\pi} \int dz\, 2\pi\delta\left(z\frac{1}{a_{Universe}} - \omega_1 - \omega_2\right) 2\pi\delta(M_H - \omega_1 - \omega_2) \\
&\quad \times \frac{e^{\frac{-8\pi\omega_1}{a_{Universe}}}}{1 - e^{\frac{-8\pi\omega_1}{a_{Universe}}}} \frac{e^{\frac{-8\pi\omega_2}{a_{Universe}}}}{1 - e^{\frac{-8\pi\omega_2}{a_{Universe}}}} \frac{1}{a_{early\ Universe}^2} \\
&\quad \times \left[\frac{z}{a_{early\ Universe}(\omega_1+\omega_2)}\left(\frac{4\Gamma\left(\frac{7}{2}\right)}{3}\right)\right]^2 \frac{\gamma_\mu \omega_1^\mu + M_b}{\omega_1{}^2 - M_b{}^2} \frac{\gamma_\nu \omega_2^\nu + M_b}{\omega_2{}^2 - M_b{}^2} \\
&\quad \times N_{color}\delta_{cd} \frac{(\omega_1+\omega_2)^2 \alpha_s}{24 v^2 M_H{}^2}\left[-4C_F \ln\left(1 - \frac{(\omega_1+\omega_2)^2}{4M_H{}^2}\right) + 2C_F \ln\left(\frac{(\omega_1+\omega_2)^2}{4M_H{}^2}\right) + 2C_F\right] \\
&\approx N_{color}\delta_{cd} \frac{M_b{}^2 \alpha_s}{24 v^2 M_H{}^2}\left[-4C_F \ln\left(1 - \frac{1}{4a_{Universe}^2 M_H{}^2}\right) + 2C_F \ln\left(\frac{1}{4}\right) + 2C_F\right] \\
&\quad \times \frac{1}{a_{early\ Universe}^2}\left[\frac{a_{Universe}}{a_{early\ Universe}}\left(\frac{4\Gamma\left(\frac{7}{2}\right)}{3}\right)\right]^2 \frac{e^{\frac{-8\pi}{a_{Universe}^2}}}{1 + e^{\frac{-8\pi}{a_{Universe}^2}}}\left[\frac{M_H^2}{M_b^2} - \ln\left(\frac{M_H - M_b}{M_H + M_b}\right)\right]
\end{aligned} \quad (36)$$

where we define the quark part, Higgs part and quark propagator for this integral as:

$$L^{Universe \to q\bar{q}} = \frac{e^{\frac{-8\pi\omega_1}{a_{Universe}}}}{1+e^{\frac{-8\pi\omega_1}{a_{Universe}}}} \frac{e^{\frac{-8\pi\omega_2}{a_{Universe}}}}{1+e^{\frac{-8\pi\omega_2}{a_{Universe}}}} \frac{1}{a_{early\ Universe}^2} \left[ \frac{z}{a_{early\ Universe}(\omega_1+\omega_2)} \left(\frac{4\Gamma(\frac{7}{2})}{3}\right) \right]^2 \quad (37)$$

$$L^{q\bar{q} \to H} = N_{color}\delta_{cd} \frac{(\omega_1+\omega_2)^2 \alpha_s}{24v^2 M_H^2} \left[ -4C_F \ln\left(1 - \frac{(\omega_1+\omega_2)^2}{4M_H^2}\right) + 2C_F \ln\left(\frac{(\omega_1+\omega_2)^2}{4M_H^2}\right) + 2C_F \right] \quad (38)$$

$$P_q = \frac{\gamma_\mu \omega_1^\mu + M_b}{\omega_1^2 - M_b^2} \quad (39)$$

$z$ is defined as the fraction of Universe mass that transforms to quark energy, $N_{color}$ is the number of colors, "c" and "d" are presented as the color indices for quarks.

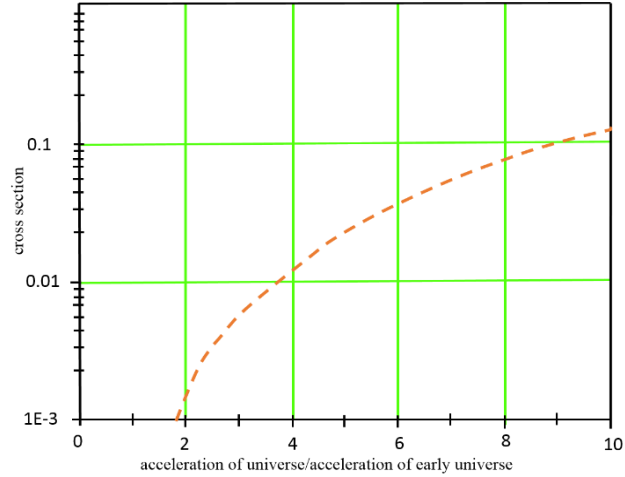

**Figure 5.** The Higgs boson production cross section, ($\sigma(Universe \to q\bar{q} \to Higgs)$), as a function of acceleration of Universe.

In Fig. 5 we present the results for the cross section of Higgs boson produced via quark-anti quark interactions near event horizon of Universe as a function of acceleration of Universe. In this plot we choose $M_b = 5\ GeV$, $M_H = 125\ GeV$, $a_{early\ Universe}$ is normalized to one and all accelerations are compared with this acceleration. As can be seen from the Fig. 5, the cross sections increases rapidly when the acceleration increases.

Now the cross section for Higgs boson which are emitted directly from Universe is studied. The Universe vacuum for Higgs will evolve into a state, called the Unruh state that can be obtained as [12]:

$$|Universe\rangle_{in \otimes out} = \frac{1}{\cosh(r_\omega)} \sum (\tanh r_\omega)^n |n\rangle_{in} \otimes |n\rangle_{out} \tag{40}$$

where $\tanh(r_\omega) = e^{\frac{-4\pi\omega}{a_{Universe}}}$ and $\cosh(r_\omega) = \left(1 - e^{\frac{-4\pi\omega}{a_{Universe}}}\right)^{-\frac{1}{2}}$.

In order to calculating Higgs cross section, it is needed to multiply its distribution by Universe production cross section. So it can be written as follow [1]:

$$\begin{aligned}
N^{Higgs}_{\omega,in} &= {}_{in \otimes out}\langle Universe|a^\dagger_{in} a_{in}|Universe\rangle_{out \otimes in} \\
&= {}_{out}\langle n|\ {}_{in}\langle n-1|\frac{1}{\cosh^2(r_\omega)} \sum_{n=0}^{\infty} \tanh^{2n}(r_\omega)\ n|n-1\rangle_{in}\ |n\rangle_{out} \\
&= {}_{out}\langle n|\ \frac{1}{\cosh^2(r_\omega)} \sum_{n=0}^{\infty} \tanh^{2n+2}(r_\omega)\ (n+1)\ |n\rangle_{out} \\
&= \frac{\sinh^2(r_\omega)}{\cosh^4(r_\omega)} \sum_{n=0}^{\omega} \tanh^{2n}(r_\omega)\ (n+1) = \frac{\sinh^2(r_\omega)}{\cosh^4(r_\omega)} \frac{1}{(1-\tanh^2(r_\omega))^2} \\
&= \sinh^2(r_\omega) \frac{\cosh^4(r_\omega)}{\cosh^4(r_\omega)} = \sinh^2(r_\omega) = \frac{e^{\frac{-8\pi\omega}{a_{Universe}}}}{1 - e^{\frac{-8\pi\omega}{a_{Universe}}}}
\end{aligned} \tag{41}$$

where $a^\dagger_{in}$ and $a_{in}$ are the creation and annihilation operators that operate on Universe inside the states of Higgs. The Higgs boson production cross section ($\sigma^3_H$) is given by:

$$\begin{aligned}
\sigma^3_{Universe \to q} &= \int d\omega\ N^{Higgs}_{\omega,in} \sigma_{Universe} \\
&\approx \frac{1}{\sigma^2_{early\ Universe}} \left[\frac{a_{Universe}}{a_{early\ Universe}} \left(\frac{4\Gamma(\frac{7}{2})}{3}\right)\right]^2 \\
&\times \frac{a_{Universe}}{8\pi} \left[\ln(a_{Universe} M_H) + \ln(a^2_{Universe} M_H^2 - 1) + 2\ln(2)\right]
\end{aligned} \tag{42}$$

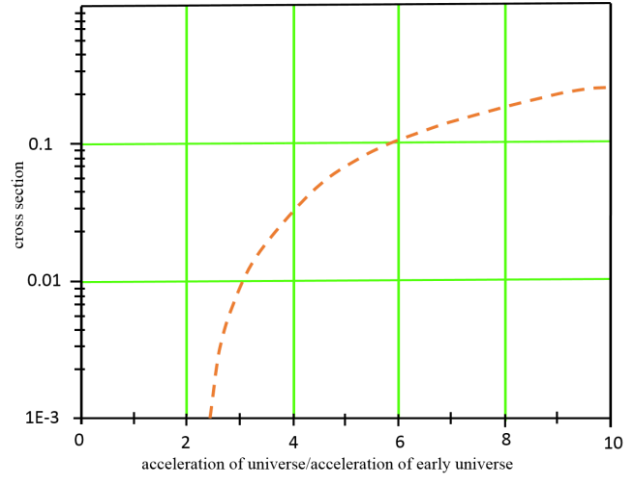

**Figure 6.** The Higgs boson production cross section, ($\sigma(Universe \rightarrow Higgs)$), as a function of acceleration of Universe

In Fig. 6 we present the cross section for Higgs boson radiated from event horizon of Universe as a function of acceleration of Universe. In this plot we choose $M_H = 125\ GeV$ for Higgs boson mass. The Higgs boson production cross section increases as the acceleration of Universe increases. Comparing Figs. (4,5,6), we observe that at lower acceleration, ($a_{Universe} < 2.5\ a_{early\ Universe}$ ), the Universe will not be able to emit Higgs, but is still able to produce a quark and eventually for $a_{Universe} < 1.5\ a_{early\ Universe}$ the Universe can only emit massless gluons. As the acceleration of Universe at the LHC increases, ($a_{Universe} > 2.5\ a_{early\ Universe}$ ), most of Higgs boson production will be due to Unruh effect near event horizon of Universe.

## 4. The Neutrino Production Due to Accelerating Universe

In this section the radiation of neutrino from event horizon of accelerating Universe is analyzed. If Universe is accelerated, many neutrinos produce due to Unruh effect near event horizon of Universe. Using the calculations in ref. [13], we have:

$$|Universe, s\rangle_{in \otimes out} = \cos(r_\omega) \sum_{n=0,1} (\tanh r_\omega)^n\ |n, s\rangle_{in} \otimes |\bar{n}, \bar{s}\rangle_{out} \qquad (43)$$

$$\tanh(r_\omega) = e^{\frac{-4\pi\omega}{a_{Universe}}} \quad , \quad \cos(r_\omega) = \left(1 + e^{\frac{-4\pi\omega}{a_{Universe}}}\right)^{-\frac{1}{2}}$$

where $|n,s\rangle_{in}$ and $|\bar{n},\bar{s}\rangle_{out}$ are neutrino and anti-neutrino states inside and outside of event horizon of accelerating Universe. The thermal distribution for this neutrino production inside the accelerating Universe is:

$$\begin{aligned} N_{\omega,in}^{neutrino} &= {}_{in\otimes out}\langle Universe, s|c_{s,in}^\dagger c_{s,in}|Universe, s\rangle_{out\otimes in} \\ &= {}_{out}\langle \bar{1},\bar{s}|\,{}_{in}\langle 0,s|\sin(r_\omega)^2|0,s\rangle_{in}|\bar{1},\bar{s}\rangle_{out} = \sin(r_\omega)^2 = \frac{e^{\frac{-8\pi\omega}{a_{Universe}}}}{1+e^{\frac{-8\pi\omega}{a_{Universe}}}} \end{aligned} \quad (44)$$

where $c_{s,in}^\dagger$ and $c_{s,in}$ are creation and annihilation operators that operate on Universe inside states of neutrino. $N_{\omega,in}^{neutrino}$ is the thermal distribution for neutrino with energy $\omega$ inside the Universe. Missing thermal distribution outside the event horizon of Universe can be calculated as follow:

$$\begin{aligned} N_{\omega,out}^{neutrino} &= {}_{in\otimes out}\langle Universe, s|c_{s,out}^\dagger c_{s,out}|Universe, s\rangle_{out\otimes in} \\ &= {}_{out}\langle 0,\bar{s}|\,{}_{in}\langle 1,s|\sin(r_\omega)^2|1,s\rangle_{in}|0,\bar{s}\rangle_{out} = \sin(r_\omega)^2 = \frac{e^{\frac{-8\pi\omega}{a_{Universe}}}}{1+e^{\frac{-8\pi\omega}{a_{Universe}}}} \end{aligned} \quad (45)$$

Where $c_{s,out}^\dagger$ and $c_{s,out}$ are creation and annihilation operators that operate on Universe outside states of neutrino. $N_{\omega,out}^{neutrino}$ is the thermal distribution for neutrinos with energy $\omega$ outside the Universe.

As mention before, in black hole problem we detect this distribution as an observer outside the event horizon of black hole. However in accelerating Universe, we miss this distribution. The total neutrino cross section near this event horizon is:

$$\begin{aligned} \sigma_{Universe\to neutrino} &= \int d\omega\, N_{\omega,in}^{neutrino} \sigma_{Universe} \\ &= \frac{1}{a_{early\,Universe}^2}\left[\frac{a_{Universe}}{a_{early\,Universe}}\left(\frac{4\Gamma\left(\frac{7}{2}\right)}{3}\right)\right]^2 \\ &\quad \times \frac{a_{Universe}}{8\pi}\left[\ln\left(1+e^{\frac{-8\pi}{a_{Universe}^2}}\right) + 2\ln(2)\right] \end{aligned} \quad (46)$$

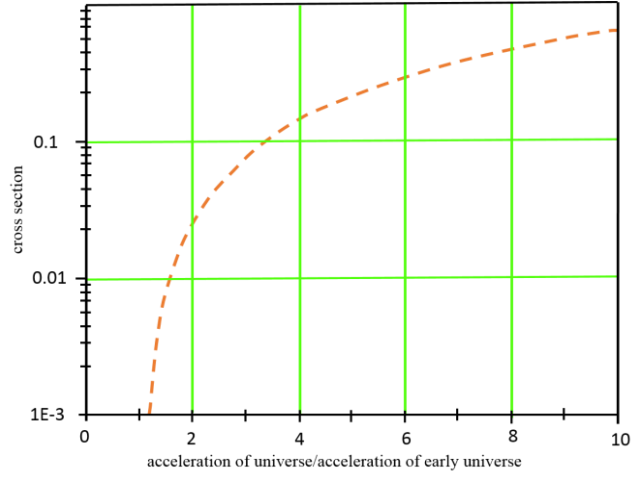

**Figure 7.** The neutrino production cross section, ($\sigma(Universe \rightarrow neutrino)$), as a function of acceleration of Universe

In Fig. 7, we show the neutrino cross section as a function of the acceleration of Universe. When the acceleration of Universe is increased, neutrino cross section becomes very large due to its ignorable mass. The detecting these particles can be a signature of acceleration of Universe.

## 5. The Production Cross Section for Dark Energy in Accelerating Universe

In this section we want to consider the dark energy states in accelerating Universe. The dark field as a massless scalar field satisfies the following wave equation:

$$\left[(-g)^{\frac{1}{2}}\frac{\partial}{\partial x^\mu}\left[g^{\mu\nu}(-g)^{\frac{1}{2}}\frac{\partial}{\partial x^\nu}\right] + V(a_{Universe})\right]\phi = 0 \tag{47}$$

where $V(a_{Universe})$ is the dark energy potential and is given by[10]:

$$V(a_{Universe}) = \frac{2}{3}c^2 M_p^2 \left[1 + \frac{b^2}{1-c^2}\right]\frac{1}{k^2 a^{\frac{3}{k}}} \tag{48}$$

In the above energy, $M_p$ is the plank mass. Requiring $\ddot{a} > 0$ for the present time, leads to $k < \frac{2}{3}$, [10,11] which can be translated into $c^2 > (1 + 3b^2)^{-1}$.

Following the calculations in ref. [12], we can write the Unruh states for dark energy as massless scalar fields as following:

$$|dark\ energy, a_{Universe}\rangle_{in\otimes out} = \frac{1}{\cosh(r_{a_{Universe}})} \sum \tanh(r_{a_{Universe}})^n |n\rangle_{in} \otimes |n\rangle_{out} \quad (49)$$

where $\tanh(r_{a_{Universe}}) = e^{\frac{-4\pi V(a_{Universe})}{a_{Universe}}}$, $\cosh(r_{a_{Universe}}) = \left(1 - e^{\frac{-4\pi V(a_{Universe})}{a_{Universe}}}\right)^{-\frac{1}{2}}$.

For calculating dark cross section we need to multiply its distribution by Universe production cross section. We have:

$$\begin{aligned}
N_{a_{Universe},in}^{dark} &= {}_{in\otimes out}\langle dark|d_{in}^\dagger d_{in}|dark\rangle_{out\otimes in} \\
&= {}_{out}\langle n|\, {}_{in}\langle n-1|\frac{1}{\cosh^2(r_{a_{Universe}})} \sum_{n=0}^\infty \tanh^{2n}(r_{a_{Universe}})\, n|n-1\rangle_{in}\, |n\rangle_{out} \\
&= {}_{out}\langle n|\, \frac{1}{\cosh^2(r_{a_{Universe}})} \sum_{n=0}^\infty \tanh^{2n+2}(r_{a_{Universe}})\, (n+1)\, |n\rangle_{out} \\
&= \frac{\sinh^2(r_{a_{Universe}})}{\cosh^4(r_{a_{Universe}})} \sum_{n=0}^\omega \tanh^{2n}(r_{a_{Universe}})(n+1) \qquad (50) \\
&= \frac{\sinh^2(r_{a_{Universe}})}{\cosh^4(r_{a_{Universe}})} \frac{1}{\left(1 - \tanh^2(r_{a_{Universe}})\right)^2} = \sinh^2(r_{a_{Universe}}) \frac{\cosh^4(r_{a_{Universe}})}{\cosh^4(r_{a_{Universe}})} \\
&= \sinh^2(r_{a_{Universe}}) = \frac{e^{\frac{-8\pi V(a_{Universe})}{a_{Universe}}}}{1 - e^{\frac{-8\pi V(a_{Universe})}{a_{Universe}}}}
\end{aligned}$$

where $d_{in}^\dagger$ and $d_{in}$ are the creation and annihilation operators that act on Universe inside the states of dark energy. We derive the dark production cross section ($\sigma_{Universe \to dark\ energy}$) and missing dark energy cross section as:

$$\begin{aligned}
\sigma_{Universe \to dark} &= \int_{V(a_{early\ Universe})}^{V(a_{Universe})} dN_{a_{Universe},in}^{dark}\, \sigma_{Universe} \\
&\approx \frac{1}{a_{early\ Universe}^2} \left[\frac{a_{Universe}}{a_{early\ Universe}} \left(\frac{4\Gamma(\frac{7}{2})}{3}\right)\right]^2 \times \frac{a_{Universe}}{8\pi} \\
&\times \left[\ln\left(\frac{a_{early\ Universe}^{\frac{3}{k}-1}}{a_{Universe}^{\frac{3}{k}-1}}\right) + \ln\left(\frac{\frac{4}{9}c^4 M_p^4 \left[1+\frac{b^2}{1-c^2}\right]^2 \frac{1}{k^4 a_{Universe}^{\frac{6}{k}-2}} - 1}{\frac{4}{9}c^4 M_p^4 \left[1+\frac{b^2}{1-c^2}\right]^2 \frac{1}{k^4 a_{early\ Universe}^{\frac{6}{k}-2}} - 1}\right) + 2\ln(2)\right]
\end{aligned} \quad (51)$$

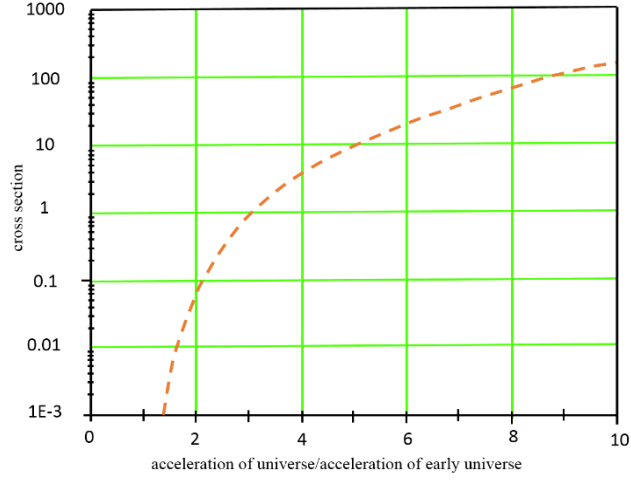

**Figure 8.** The dark energy production cross section, ($\sigma(Universe \to dark\ energy)$), as a function of acceleration of Universe

The thermal distribution for missing dark energy can be calculated as following:

$$\begin{aligned}
N^{dark}_{a_{Universe},out} &= {}_{in\otimes out}\langle dark|d^{\dagger}_{out}d_{out}|dark\rangle_{out\otimes in} \\
&= {}_{out}\langle n-1|\,{}_{in}\langle n|\frac{1}{cosh^2(r_{a_{Universe}})}\sum_{n=0}^{\infty} tanh^{2n}(r_{a_{Universe}})\,n|n\rangle_{in}\,|n-1\rangle_{out} \\
&= {}_{in}\langle n|\frac{1}{cosh^2(r_{a_{Universe}})}\sum_{n=0}^{\infty} tanh^{2n+2}(r_{a_{Universe}})\,(n+1)\,|n\rangle_{in} \\
&= \frac{sinh^2(r_{a_{Universe}})}{cosh^4(r_{a_{Universe}})}\sum_{n=0}^{\omega} tanh^{2n}(r_{a_{Universe}})\,(n+1) \qquad (52) \\
&= \frac{sinh^2(r_{a_{Universe}})}{cosh^4(r_{a_{Universe}})}\frac{1}{\left(1-tanh^2(r_{a_{Universe}})\right)^2} = sinh^2(r_{a_{Universe}})\frac{cosh^4(r_{a_{Universe}})}{cosh^4(r_{a_{Universe}})} \\
&= sinh^2(r_{a_{Universe}}) = \frac{e^{\frac{-8\pi V(a_{Universe})}{a_{Universe}}}}{1-e^{\frac{-8\pi V(a_{Universe})}{a_{Universe}}}}
\end{aligned}$$

and the missing cross section is given by:

$$\sigma_{missing\ Universe\to dark\ energy} = \int_{V(a_{early\ Universe})}^{V(a_{Universe})} dN^{dark}_{a_{Universe},out}\,\sigma_{Universe} \qquad (53)$$
$$\approx \sigma_{Universe\to dark\ energy}$$

In Fig. 8 we present the cross section for dark energy as a function of acceleration of Universe. Here we take $M_P = 1\ TeV, K = 0.3, b^2 = 0.1, c^2 = 0.9$. Also we normalize the early acceleration to one and compare other accelerations of Universe with it. Comparing this figure with Figs. (2-8), we find that the cross section for dark energy is dominated by QCD matter, Higgs boson and neutrino. This result is consistent with previous calculation for dark energy.

## 6. Summary

In this research, the production cross sections for quark, gluon, neutrino and dark energy due to acceleration of Universe are investigated. To obtain these cross sections, the Universe production cross section is multiplied by the particle or dark energy distribution in accelerating Universe. Also the missing cross sections for each matter and dark energy due to formation of event horizon is discussed. Then different channels for producing Higgs boson in accelerating Universe are studied and the cross section in each channel is obtained. Finally the production cross section for dark energy is compared with particle cross sections and it is found that the cross section for dark energy is dominated by QCD matter, Higgs boson and neutrino. This result is consistent with previous predictions for dark energy cross section.

**References:**

[1] A. Sepehri et al., Can. J. Phys. 90: 25–37 (2012)

[2] A. Sepehri, M.R. Setare, Journal of High Energy Physics, 1503 (2015), 079 (1-17)

[3] S. Perlmutter et al., Supernova Cosmology Project, Astrophys. J. 517, 565 (1998)

[4] A. Sepehri, et al., Astrophysics and Space Science, Volume 344, Issue 1, 2013, Pages 79-86

[5] I. Fuentes-Schuller and R. B. Mann, Phys. Rev. Lett. 95, 120404 (2005).

[6] P. M. Alsing, I. Fuentes-Schuller, R. B. Mann, and T. E. Tessier, Phys. Rev. A74, 032326 (2006).

[7] D.N. Spergel, et al., WMAP Collaboration, Astrophys. J. Suppl. 148 (2003) 175

[8] M. Tegmark, et al., SDSS Collaboration, Phys. Rev. D 69 (2004) 103501

[9] E.J. Copeland, M. Sami, S. Tsujikawa, Int. J. Mod. Phys. D 15 (2006)

[10] L. N. Granda and A. Oliveros, Phys. Lett. B 671, 199(2009).

[11] Ahmad Sheykhi, Phys. Rev. D.84, 107302 (2011)

[12] Doyeol Ahn, Phys. Rev. D74: 084010, (2006). W. G. Unruh, *Phys. Rev. D* **14**, 870 (1976).

[13] D. Ahn, Y.H. Moon, R. B. Mann, I. Fuentes-Schuller, JHEP0806:062, (2008).

[14] C. Itzykson and F-B. Zuber, Quantum Field Theory, McGraw-Hill, New York (1980).

[15] Jun-Chen Su, J.Phys. G27 (2001) 1493-1500

[16] N. D. Birrell and P. C. W. Davies, *Quantum fields in curved space* (Cambridge University Press, New York, 1982).

[17] R. M. Wald, Quantum field theory in curved spacetime and black hole thermodynamics (The University of Chicago Press, Chicago, 1994).

[18] S. Dimopoulos and G. Landsberg, Phys. Rev. Lett. 87, 161602 (2001)

[19] S. B. Giddings and S. Thomas, Phys. Rev. D 65, 056010 (2002)

[20] D. G. Boulware, Phys. Re*v*. D **12**, 350 (1975).

[21] Pushpalatha C. Bhat, Russell Gilmartin, Harrison B. Prosper, Phys.Rev. D62 (2000) 074022

[22] M. Spira, A. Djouadi, D. Graudenz and P. M. Zerwas, Nucl. Phys. B 453 (1995) 17

[23] Nikolaos Kidonakis, Phys. Rev. D 77:053008, (2008). Mathias Butenschoen, Frank Fugel, Bernd A. Kniehl, Phys. Rev. Lett. 98:071602, (2007).